# Using Cluster Curves to Control Software Development Projects

Jürgen Münch[a], Jens Heidrich[b]
[a]*Fraunhofer Institute for Experimental Software Engineering, Kaiserslautern, Germany*
[b]*University of Kaiserslautern, Kaiserslautern, Germany*
muench@iese.fhg.de, heidrich@informatik.uni-kl.de

**Abstract**

*Online interpretation and visualization of project data are gaining increasing importance on the long road towards predictable and controllable software project execution. This paper sketches the Sprint 1 controlling approach for software development projects and gives first evaluation results. The approach is grounded on the usage of context-oriented cluster curves and integrated in the framework of software project control centers.*

## 1. Introduction

One means to institutionalize measurement-based software development on the basis of explicit models is the development and establishment of so-called software project control centers (SPCC) for systematic quality assurance and management support. An SPCC is comparable to a control room, which is a well-known term in the mechanical production domain. We define a software project control center as a means for collecting, interpreting, and visualizing measurement data in order to provide purpose- and role-oriented information to all parties involved (e.g., project manager, quality assurer) during the execution of a project. This includes, for instance, monitoring defect profiles, detecting abnormal effort deviations, cost estimation, and cause analysis of plan deviations. This paper describes a special controlling technique for an SPCC called Sprint 1. The technique is based on cluster analysis introduced by Li and Zelkowitz in 1993 [1] and identification of trend changes introduced by Tesoriero and Zelkowitz in 1998 [2]. We see the following advantages of the technique: The prediction and control of project progression is directly based on experience with past projects (or experiments), the accuracy of planned curves is increased by context- and data-driven selection of cluster curves, the approach is directly applicable without a number of reference applications, the adaptation of planned curves takes place dynamically, and, finally, the storage and management of experience curves is well-structured (structuring through context vector approach and clustering approach). Thus, the technique is tailored for the specifics of software development (e.g., context-dependability of development processes).

## 2. Technique

The prerequisite for a successful application of Sprint 1 is that a software development organization has already performed a number of similar projects and measured at least one key attribute (e.g., effort per development phase) for each of these projects. Additionally, the context for each of these projects (i.e., the boundary conditions such as organizational, personal and technical constraints) needs to be characterized. The technique can be sketched as follows: First, the context-specific measurement data from former projects is analyzed in order to identify clusters. Based on the context of the project to be controlled, the technique selects a suitable cluster and uses its cluster curve (mean of all curves within a cluster) for predicting the attributes to be controlled. During the enactment of the project, the prediction is adapted based on actual project data. This leads to an empirical-based prediction and to flexibility for project and context changes.

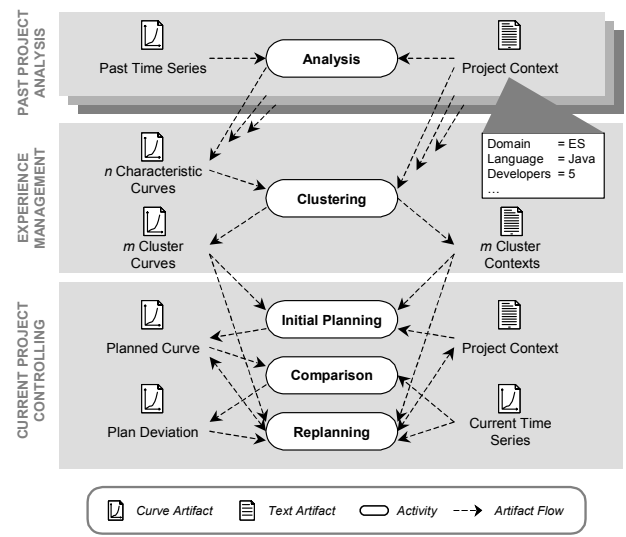

**Figure 1.** Artifact flow of the approach.

The proposed controlling technique basically consists of five steps (see Figure 1): (1) *Analysis*. The first step of the technique analyzes the time-series of the measured attributes per completed project in order to get so called characteristic curves of the attributes considered. To every characteristic curve a corresponding context is assigned, which represents the project environment that the curve originates from. A context description comprises all factors with a proven or assumed impact on the attribute values (such as people factors, technology factors, organizational factors, process factors). (2) *Clustering*. For a certain attribute clustering is used to identify groups of characteristic curves that belong together. These are, for instance, curves, whose distance is less than a certain threshold. All characteristic curves within one cluster are averaged to get a model (the so-called cluster curve), which stands for the whole cluster. Again, a context is assigned to the aggregated curves based on a similarity analysis of the project contexts of the cluster. (3) *Initial Planning*. During project planning, the relevant project attributes are estimated on the basis of cluster curves for the respective attributes of previous projects. At the beginning of the project no actual data are available and therefore, context characteristics must be used in order to find a suitable cluster curve for the project attribute under consideration (see Figure 1). (4) *Comparison*. During the enactment of the project the current values of an attribute are compared with the predicted values of the cluster curve. If a plan deviation occurs, the predicted values have to be adapted with respect to the new project situation. Therefore, the distance between the two curves has to be computed regularly. If the distance is above (or below) a most tolerable threshold, project management has to be informed in order to initiate dynamic replanning steps, and a new cluster curve has to be sought in order to make a new prediction. (5) *Replanning*. In case of significant plan deviation, the causes for the deviation have to be determined. We basically distinguish three different cases: The first one is that the experience we used to build up our prediction model was wrong. The second one is that the characteristics we assumed for our project were wrong (e.g., the experience of the developers was low instead of high). In this case we have to adapt the project context. The third case is that problems occur in the project that lead to a change of the characteristics of the project (e.g., technology changed). In all three cases we can try to identify a new cluster curve within the set of computed clusters. Basically, there are three ways to choose a suitable cluster for prediction: The first one is matching the contexts of the actual project and the cluster curves (like step 3). The second possibility is to use the current data of a certain attribute, which has been measured during the enactment of the project up to the present, and match it with the cluster curves in order to find the best cluster curve for prediction. This is a dynamic assignment approach, which incorporates actual project behavior. The third option is to combine the static and dynamic approach to get a hybrid one. If both possibilities lead to different clusters, a set of exception handling strategies can be applied and reasons be sought. Afterwards, step 4 is iterated and uses the adapted prediction in order to further control the project.

## 3. Case Study Results

The technique has been initially evaluated using data from 25 projects. Clustering has been performed on 17 randomly selected projects using 10 context parameters. 4 projects were used for testing the controlling technique. 5 clusters could be identified. The determination of characteristic cluster contexts was based on majority decision and averaging (in case of numerical context characteristics). The selection of the appropriate cluster curves was only based on context similarity, i.e., the static selection approach was used. The application of the controlling technique on the 4 test projects showed that the context-based selection of cluster curves led to curves with small average deviations between predicted and actual values. We expect that the promising results can be further improved, if one adds the actual data for determining the prediction curve.

## 4. Related Work and Conclusion

Implementations of SPCCs are introduced by the following approaches: Provence, Amadeus, Ginger2, SME (Software Management Environment), WebME, and PAMPA. All these approaches reside in the software development domain, but approaches from the business or production process domain can also be found in this field. There exist a number of controlling techniques in the context of the listed approaches with different emphases. Most of them are not based on empirical data of a specific organization, but rather rely on very generic mathematical models. In contrast, the technique proposed here is based on organizational experience and allows for individual context adaptation. The use of dynamic simulation modeling techniques (such as System Dynamics) for prediction could be a good complementation of the technique in the case that not enough empirical data is available.